\documentclass{ws-mpla}

\usepackage{graphicx} 

\newcommand{\ltapprox}{\raisebox{-0.5ex}{$\,\stackrel{<}{\scriptstyle
\sim}\,$}}
\newcommand{\ri}{r_{\rm i}}
\newcommand{\rw}{r_{\rm w}}
\newcommand{\rg}{r_{\rm g}}
\newcommand{\phiG}{\phi_{\scriptscriptstyle \rm G}}
\newcommand{\cs}{c_{\rm s}}

\newcommand{\OmegaK}{\Omega_{\scriptscriptstyle \rm K}}
\newcommand{\fa}{f_{\rm a}}
\newcommand{\fw}{f_{\rm w}}
\newcommand{\Fd}{F_{\rm d}}
\newcommand{\Ld}{L_{\rm d}}

\newcommand{\Pa}{P_{\rm a}}
\newcommand{\Pw}{P_{\rm w}}

\newcommand{\Mdot}{\dot M}
\newcommand{\Mdota}{\dot M_{\rm a}}
\newcommand{\Mdotw}{\dot M_{\rm w}}
\newcommand{\ddr}{\frac{\rm d}{{\rm d}r}}
\newcommand{\dr}{{\rm d}r}
\newcommand{\dz}{{\rm d}z}

\begin{document}

\markboth{Kuncic \& Bicknell}
{A New Standard Accretion Disk Theory}

\catchline{}{}{}{}{}

\title{
TOWARDS A NEW STANDARD THEORY FOR \\
ASTROPHYSICAL DISK ACCRETION
}

\author{\footnotesize ZDENKA KUNCIC}

\address{School of Physics, University of Sydney\\
Sydney, NSW, 2006, Australia \\
Z.Kuncic@physics.usyd.edu.au}

\author{GEOFFREY V. BICKNELL}

\address{Research School of Astronomy and Astrophysics,
Australian National University \\
Canberra, ACT, 2611, Australia \\
Geoff.Bicknell@anu.edu.au
}

\maketitle

\pub{Received (Day Month Year)}{Revised (Day Month Year)}

\begin{abstract}
We briefly review recent developments in black hole accretion disk
theory, placing new emphasis on the vital role played by magnetohydrodynamic (MHD)
stresses in transporting angular momentum. The apparent universality
of accretion-related outflow phenomena is a strong indicator that vertical transport of angular momentum by large-scale MHD torques is important and may even dominate radial transport by small-scale MHD turbulence.
This leads to an enhanced overall rate of angular momentum
transport and allows accretion of matter to proceed at an interesting
rate. Furthermore, we argue that when vertical transport is important,
the radial structure of the accretion disk is modified and this affects the disk emission spectrum. We present a simple
model demonstrating that energetic, magnetically-driven outflows give
rise to a disk spectrum that is dimmer and redder than a standard accretion disk accreting at the same rate. We
briefly discuss the implications of this key result for accreting
black holes in different astrophysical systems. 
\end{abstract}

\ccode{PACS Nos.: include PACS Nos.}

\section{Introduction}	

Accretion disks are ubiquitous in astrophysics.
They form around newly born stars, in stellar binary systems and
around supermassive black holes in the nuclei of galaxies.
Their efficiency at extracting gravitational energy is truly remarkable:
in the case of a rotating black hole, the conversion efficiency can
exceed a staggering 40\% of the equivalent rest mass energy of the accreting
matter.
It is now widely accepted that the most luminous
sources in the Universe, namely quasars and gamma-ray bursts, must
necessarily be powered by accretion onto a black hole.
It is also widely accepted that high-energy astrophysical sources,
such as X-ray binaries and Active Galactic Nuclei (AGN), are also
accretion-powered.

The standard theory of astrophysical disk accretion was formulated over
thirty years ago. \cite{prinrees72,novthorn73,SS73}
Since then, arguably the most important theoretical milestone has been
the demonstration via numerical simulations that the removal of angular
momentum required for accretion to proceed is mediated by 
magnetohydrodynamic (MHD) turbulence driven by the weak-field
magnetorotational instability (MRI -- see Ref.~\refcite{balbus03} for a review).
Notwithstanding these groundbreaking developments, it still remains
unclear precisely how energy can be channelled from the accretion flow
to an external diffuse region where high-energy processes can operate.
This outstanding issue is inextricably linked with the  ubiquity of outflow phenomena associated with accretion disks.
Indeed, the formation and nature of relativistic jets remains one of the
most formidable theoretical problems in this field.

In this brief review, we highlight some recent theoretical progress made towards
understanding vertical transport in accretion disks. \cite{kunbick04}
A macroscopic, mean field approach is adopted for the MHD turbulence.
This allows us to identify the dominant mechanisms responsible for
angular momentum transport as well as the main contributions to the global
energy budget of the system. We also demonstrate quantitatively how outflows, both
Poynting-flux-dominated and mass-flux-dominated, can modify the
spectrum of disk emission substantially from that predicted by
standard accretion disk theory.

\section{MHD Disk Accretion}

In the following, we present statistically-averaged equations in a
cylindrical $(r,\phi,z)$ coordinate system for a fluid which is
time-independent and axisymmetric in the mean
($\langle \partial / \partial t \rangle = \langle \partial /\partial
\phi \rangle =0$, where the angle brackets denote ensemble-averaged quantities).
We use Newtonian physics throughout, with the gravitational
potential $\phiG = -GM(r^2 + z^2)^{-1/2}$.
The fluid description is valid down to the innermost marginally stable 
orbit $\ri \approx 6\rg$, where $\rg = GM/c^2$ is the gravitational
radius of a black hole of mass $M$.
The mean-field conservation equations are integrated vertically over an
arbitrary disk scaleheight, $h = h(r)$.
Quantities calculated at the disk surface ($z= \pm h$) are denoted by a $\pm$
superscript, $X^{\pm}$, and we assume reflection symmetry about the disk midplane,
so that $|X^+| = |X^-|$.
Midplane values of variables are denoted by $X_0$.
We assume $h$ is much less than the radius (the ``thin disk''
approximation), so that quantities of order $h/r$ and ${\rm d}h/{\rm d}r$ are neglected.

We adopt a mass-weighted statistical averaging approach in 
which all variables are decomposed into mean and fluctuating parts,
with intensive variables such as the velocity mass averaged according
to
\begin{equation}
v_i = \tilde v_i + v_i^\prime \qquad , \qquad \langle \rho v_i^\prime
\rangle =0 \qquad ,
\end{equation}
while extensive variables, such as density, pressure and magnetic
field, averaged the following way:
\begin{eqnarray}
\rho = \bar \rho + \rho^\prime \> &,& \> \langle \rho^\prime
\rangle =0 \qquad ,  \\
p = \bar p + p^\prime \> &,& \>  \langle p^\prime
\rangle =0 \qquad , \\
B_i = \bar B_i + B_i^\prime \> &,&  \> \langle B_i^\prime
\rangle =0 \qquad .
\end{eqnarray}
Note that intensive averages are denoted by a tilde, extensive
averages are denoted by a bar and fluctuating components are denoted
by a prime. The fluctuating velocity components are restricted
to subsonic speeds because the MRI is a weak-field instability which
drives subsonic turbulence, with $ \langle \rho v^{\prime 2} \rangle
\ltapprox \langle \rho c_{\rm s}^2 \rangle \ll \rho r^2 {\OmegaK}^2$, where
$c_{\rm s} = (kT/\mu m_{\rm p})^{1/2}$
is the local sound speed and $\OmegaK = (GM/r^3)^{1/2}$ is the keplerian angular
velocity. The only restriction we place on the mean fluid velocity
components is that they
satisfy $\tilde v_\phi \gg \tilde v_r ,  \tilde v_z$ and that $\tilde
v_r$ and $\tilde v_\phi$ are
independent of $z$, simplifying the vertical integration.
The combined magnetic and Reynolds stresses are defined by
\begin{equation}
\langle t_{ij} \rangle = \langle \frac{B_i B_j}{4\pi} \rangle -
\delta_{ij} \langle \frac{B^2}{8\pi} \rangle - \langle \rho v_i^\prime v_j^\prime \rangle
\label{e:tij}
\end{equation}

In what follows, we summarize the salient equations for radial and
vertical transport of mass, angular momentum and energy. A full
derivation of these equations can be found in our earlier
paper. \cite{kunbick04} For clarity, we have simplified the
presentation of the relevant equations by omitting negligible
correlation terms and by removing the notation for averaged extensive
and intensive quantities. Thus, averaged quantities are implicitly assumed.

\subsection{Mass transfer}
\label{s:mass}

Vertical integration of the mean-field continuity equation gives
\begin{equation}
\ddr \int_{-h}^{+h} 2 \pi r  \rho  v_r \> \dz
\, + \, 4 \pi r  \rho^+  v_z^+  = 0  \qquad .
\label{e:mass_int}
\end{equation}
We now introduce the usual definitions for the surface mass density,
\begin{equation}
\Sigma(r) \equiv \int_{-h}^{+h}  \rho \> \dz
\label{e:sigma_defn}
\end{equation}
and mass accretion rate,
\begin{equation}
\Mdota (r) \equiv 2\pi r  \Sigma (-  v_r)  \qquad .
\label{e:Mdota}
\end{equation}
We also introduce an analogous mass outflow rate,
\begin{equation}
\Mdotw (r) =  \int_{r}^{\infty} 4\pi r  \rho^+  v_z ^+{\rm d}r
\label{e:Mdotw}
\end{equation}
associated with a mean vertical velocity $ v_z^+$ at the disk 
surface, i.e. at the base of a disk wind. This is the total rate at which mass leaves the disk surface at radii exterior to $r$.

In terms of the above definitions, the vertically integrated continuity equation,
(\ref{e:mass_int}), can be written as
\begin{equation}
\ddr \Mdota (r) = 4\pi r  \rho^+ v_z ^+ = - \ddr \Mdotw (r)
\label{e:dM}
\end{equation}
implying that
\begin{equation}
\Mdota (r) + \Mdotw (r) = \Mdota (\ri) + \Mdotw (\ri) = \hbox{constant} = \Mdot
\qquad ,
\label{e:Mdot}
\end{equation}
where $\Mdota (\ri) + \Mdotw (\ri)$ is the total mass flux at the innermost
stable orbit, $\ri$.
Equation (\ref{e:dM}) implies that under steady--state conditions,
the radial mass inflow decreases towards small $r$ at the same rate as the
vertical mass outflow increases in order to maintain a constant nett mass flux,
$\Mdot$, which is the nett accretion rate at $r =
\infty$, i.e. $\Mdot = \Mdota (\infty)$.

\subsection{Angular momentum}
\label{s:ang_mom}

The azimuthal component of the momentum equation is:
\begin{equation}
\frac{1}{r^2} \frac{\partial}{\partial r} \left(
r^2  \rho  v_r  v_\phi \right)
+ \frac {\partial}{\partial z} \left( \rho v_\phi v_z \right) = \frac{1}{r^2}
\frac{\partial}{\partial r}\left( r^2 \langle t_{r \phi} \rangle  \right)  
+ \frac{\partial \langle t_{\phi z} \rangle}{\partial z} \qquad .
\label{e:azimuthal}
\end{equation}
Integrating this equation  over $z$ and applying mass continuity (\ref{e:dM})
gives
\begin{equation}
\ddr \left[ \Mdota r^2 \Omega + 2\pi r^2 T_{r\phi} \right]
= r^2 \Omega \frac{{\rm d}\Mdota}{{\rm d}r}
- 4\pi r^2 \langle t_{\phi z} \rangle^+   \qquad ,
\label{e:angmom}
\end{equation}
where
\begin{equation}
T_{r \phi} = \int_{-h}^{+h} \langle t_{r \phi} \rangle \> \dz
\label{e:int_stress}
\end{equation}
is the vertically integrated $r\phi$ stress. The terms on the the left
hand side of eqn.~(\ref{e:angmom}) describe radial transport of
angular momentum associated with radial inflow (accretion) and MHD
stresses acting over the disk height. The terms on
the right hand side describe vertical transport of angular momentum
resulting from mass loss in a wind and MHD stresses on the disk
surface. The magnetic part of the MHD stresses are given by
$\langle t_{r\phi} \rangle \sim \langle B_r
B_\phi \rangle/4\pi$ and  $\langle t_{\phi z} \rangle \sim \langle
B_\phi B_z \rangle/4\pi$ (c.f. eqn.~\ref{e:tij}) .

Radially integrating  equation (\ref{e:angmom}) (taking $\Omega \approx \OmegaK$) gives
\begin{equation}
\Mdota   r^2 \Omega 
- \Mdota (\ri)   \ri^2 \Omega (\ri) =
-2\pi r^2 T_{r \phi} + 2\pi \ri^2 T_{r \phi}(\ri)  +
 \int_{\ri}^{r} \left[ \,  r^2 \Omega \frac{{\rm d}\Mdota}{{\rm d}r}
- 4 \pi r^2  \langle t_{\phi z}\rangle^+ \,  \right] \> {\rm d}r  \> .
\label{e:angmom_int}
\end{equation}
This is a generalized conservation equation for angular momentum
in accretion disks. It can be equivalently expressed in terms of the
flux of angular momentum, $\dot J$, as follows:
\begin{equation}
\dot J_{\rm a}(r) - \dot J_{\rm a} (\ri) = \dot J_r (r) - \dot J_r
(\ri) + \dot J_z (r)
\qquad ,
\label{e:Jdot}
\end{equation}
where $\dot J_{\rm a} = \Mdota   r^2 \Omega$ is the angular momentum
flux of the accreting matter, $\dot J_r = -2\pi r^2 T_{r \phi}$ is the
radial flux of angular momentum carried by the $r\phi$ stresses and $\dot
J_z = \int_{\ri}^{r} \left[ \,  r^2 \Omega \, {\rm d}\Mdota / {\rm d}r
- 4 \pi r^2  \langle t_{\phi z}\rangle^+ \,  \right] \; {\rm d}r$ is
the vertical flux of angular momentum carried by outflowing matter
and by the $\phi z$ stresses. Thus, both radial and vertical transport
can contribute to the nett rate at which matter loses angular momentum and moves
radially inwards.  In other words, the mass accretion rate, $\Mdota$,
is determined by the nett rate at which angular momentum is transported
radially outwards by MHD stresses acting over the disk height and
vertically outwards by  MHD stresses acting over the disk surface.

The most powerful astrophysical sources are inferred to be accreting matter at very high rates.
Quasars, for instance, radiate energy at an astonishing rate, typically $10^{45-48}\,{\rm erg \, s^{-1}}$. Much of this luminosity can be identified with thermal emission from an accretion disk around a supermassive  black hole. If all the extracted gravitational binding energy of the accreting matter is converted into disk radiation, then the radiative efficiency is just equal to the accretion efficiency, $\eta$, which is typically $10$\,\%. Thus, the rate at which matter needs to be accreting is $\Mdota = L/(\eta c^2) \simeq (0.2-200) \, M_\odot \, {\rm yr}^{-1}$. Identifying the dominant mechanisms responsible for such high mass accretion rates has been a
longstanding challenge confronting accretion disk theory. This problem is made even more severe  if one considers in addition conversion of energy into non-radiative forms, such as relativistic jets, which can carry a total kinetic power at least as comparable to the radiative disk luminosity.

It is interesting to compare the rate of angular momentum transport
indicated by (\ref{e:angmom_int}) with that predicted by standard
accretion disk theory. \cite{SS73} In a standard disk, angular
momentum is transported radially outwards only and the stresses
responsible for this are parameterized by a dimensionless parameter
$\alpha$ such that
\begin{equation}
T_{r \phi} = \alpha \cs h \Sigma r  \frac{\partial
  \Omega}{\partial r} 
  \label{e:alpha}
\end{equation}
with $0 \ltapprox \alpha \ltapprox 1$ for stresses due to subsonic turbulence. In addition, the stresses are assumed to vanish at $\ri$ and $\Mdota$
is necessarily constant with $r$ (since there is no vertical mass flux). Thus, the angular momentum  equation, (\ref{e:angmom_int}), reduces to
\begin{equation}
\Mdota r^2 \Omega \left[ 1 - \left( \frac{r}{\ri} \right)^{-1/2} \right] = 2\pi \, r^3 \, \alpha \, \cs \, h \Sigma \,
\left| \frac{\partial \Omega}{\partial r}  \right|
\label{e:Mdot_alpha}
\end{equation}
There are two important limitations to the $\alpha$-disk
formalism. Firstly, $\alpha$ is treated as a global parameter, so that it remains constant at all radii; there is no \textit{a priori} reason why the radial distribution of  turbulent stresses should be parametrized in this way. Secondly, because the $\alpha-$disk model remains a phenomenological
prescription, it is unclear whether the values of $\alpha$ needed to account for the inferred accretion rates in different sources correspond to a physically plausible realization of the
actual turbulent stresses in real accretion disks.

Numerical simulations may be uniquely capable of addressing these
issues. Simulations of accretion disks have demonstrated conclusively
that turbulent MHD stresses can indeed remove angular momentum from
matter, thus facilitating the accretion process (see
Ref.~\refcite{balbus03} for a review). However, 3D simulations of
MRI-generated MHD turbulence in accretion disks
have so far been unable to produce high mass accretion
rates. \cite{hawley00,stonepring01,hawley01,hawbalb02}  On the other hand, high
accretion rates are recovered from 3D MHD simulations when a
large-scale, poloidal mean magnetic field is explicitly
included. \cite{steinhenn01,kigshib05} These simulations and
others \cite{salm07}, as well as semi-analytic models \cite{campbell03},
show that at small radii, vertical transport of angular momentum by a large-scale
magnetic torque is more efficient than radial transport by MHD turbulence.

The numerical results indicate that MHD turbulence, whilst important
for the microphysics of accretion disks, cannot be solely responsible for
the efficacy of angular momentum transport. Indeed, as we show
explicitly in eqn.~(\ref{e:angmom_int}), angular momentum can be removed
vertically from an accretion disk by the mean fluid and magnetic fields
(i.e. nonzero $v_z$ and $\langle B_\phi B_z \rangle$ components at the disk
surface -- see also Ref.~\refcite{konpud00}). We can make a direct
comparison between the value of $\Mdota$ predicted by the standard
model (which considers only outward radial transport) and that
predicted by our model (which includes outward vertical transport). Using the
$\alpha$-disk formalism, the solution for
$\Mdota$ generalizes to
\begin{equation}
\Mdota = 2\pi \alpha \cs h \Sigma
\frac{r}{\Omega} \left| \frac{\partial \Omega}{\partial r}  \right|
\left( 1 + \frac{\dot J_z}{\dot J_r} \right)
\left[ 1 - \left( \frac{r}{\ri} \right)^{-1/2} \right] ^{-1}
\end{equation}
which is larger  than that predicted by the standard disk model, given
by (\ref{e:Mdot_alpha}), by a factor $1 + \dot J_z / \dot J_r$. That is, our
model predicts that in regions where the vertical angular momentum
flux exceeds the radial angular momentum flux, the local rate of mass
accretion is larger than that predicted by standard accretion disk
theory. 

We conclude here that it is large-scale, organized MHD stresses
in the form of magnetic torques and mass outflows, rather than small-scale,
turbulent MHD stresses, that lead to an enhanced \textit{overall} rate of angular
momentum transport (and hence, high $\Mdota$)  in astrophysical accretion disks. The fundamental relationship
between angular momentum and mechanical energy (${\rm d} E = \Omega {\rm d} J$)
then also implies that large-scale MHD effects must also be
responsible for driving high-energy phenomena associated with accreting
sources which must necessary originate outside the disk in a
relativistic jet or magnetized atmosphere (corona).
In the following section, we examine the global energy budget of
accretion disks in which angular momentum transport is prescribed by
the generalized conservation equation (\ref{e:angmom_int}).

\subsection{Radiative disk flux}
\label{s:energy_budget}

We now consider energy conservation in MHD disk accretion. Again, we
refer the reader to our earlier paper (Ref.~\refcite{kunbick04})
for details of the derivations.

Accretion power is the rate at which gravitational binding energy is extracted from the
accreting matter. This energy can be converted into mechanical
(e.g. kinetic, Poynting flux) and
non-mechanical (e.g. radiative) forms. The rate at which this
occurs is determined by keplerian shear in the bulk flow, $s_{r\phi} =
\frac{1}{2} r \partial \Omega / \partial r$, with $\partial \Omega /
\partial r = - \frac{3}{2} \Omega / r$. The rate per unit disk
surface area at which energy is emitted in the form of electromagnetic
radiation is determined by the internal energy equation. If there are
negligible changes in the internal energy and enthalpy of the gas,
then the disk radiative flux, $F_{\rm d}$, is approximately equal to
the rate of stochastic viscous dissipation of the turbulent energy,
which occurs on the smallest scales, either at the end of a turbulent
cascade or via magnetic reconnection. If there is negligible transport of turbulent energy from the
source region, then turbulent energy is locally dissipated at a rate
equivalent to its production rate, $\langle t_{ij} \rangle s_{ij}
\approx \langle t_{r\phi} \rangle s_{r\phi}$ (see
Ref.~\refcite{kunbick04} for a discussion on the relative importance
of production, transport and dissipation of turbulent energy). The internal energy
equation then implies
\begin{equation}
\frac{\partial F_{\rm d}}{\partial z} \approx \langle t_{r\phi} \rangle s_{r
  \phi}
  \label{e:internalE}
\end{equation}
Vertically integrating over the disk height yields the following
expression for the radiative flux emerging from the disk surface:
\begin{equation}
F_{\rm d}^+ \, \approx \, \frac{1}{2} T_{r\phi}  r  \frac{\partial
  \Omega}{\partial r}  \, = \, -\frac{3}{4} T_{r\phi} \Omega
\label{e:Fd_internal}
\end{equation}
Note that in the $\alpha$-disk formalism, defined by (\ref{e:alpha}),
$F_{\rm d}^+$ is directly proportional to $\alpha$. In a standard
disk, $\alpha$ is also directly proportional to $\Mdota$
(c.f. \ref{e:Mdot_alpha}). That is, a standard $\alpha$-disk model
predicts that the disk surface brightness varies linearly with
the mass accretion rate. Our model, in contrast, predicts that disks
can be accreting at a relatively high rate, but locally emit relatively little
radiation if vertical mean-field transport is locally more
efficient than radial turbulent transport.

According to our model,
the level of the turbulent MHD stresses $T_{r\phi}$ available for
internal dissipation is determined by the relative efficiency with
which the extracted accretion energy is converted into
non-radiative (mechanical) form. In other words, $T_{r\phi}$ is specified by the
angular momentum conservation relation (\ref{e:angmom_int}), which
gives
\begin{eqnarray}
-T_{r\phi} (r) &=& \frac{\Mdota \Omega}{2\pi } \left[ 1 - \frac{\Mdota
    (\ri)}{\Mdota (r)} \left( \frac{\ri}{r} \right)^{1/2} \right] -
    \left( \frac{\ri}{r} \right)^2 T_{r\phi} (\ri) \nonumber \\
&-& \frac{1}{2\pi r^2}
\int_{\ri}^{r} \left[ \,   r^2 \Omega  \frac{{\rm d}\Mdota}{{\rm d}r}
- 4 \pi r^2  \langle t_{\phi z}\rangle^+ \,  \right] \> {\rm d}r
\end{eqnarray}
Substituting this expression into the internal energy equation (\ref{e:Fd_internal}) yields the
following solution for the disk radiative flux:
\begin{eqnarray}
F_{\rm d}^+ (r) &\approx& \frac{3GM\Mdota (r)}{8\pi r^3} \left[ 1 - \frac{\Mdota
    (\ri)}{\Mdota (r)} \left( \frac{\ri}{r} \right)^{1/2} \right] -
    \frac{3}{4} \left( \frac{\ri}{r} \right)^2 T_{r\phi} (\ri) \Omega
    \nonumber \\
&-& \frac{3\Omega}{8\pi r^2} \int_{\ri}^{r} \left[ \,  r^2 \Omega \frac{{\rm d}\Mdota}{{\rm d}r}
- 4 \pi r^2  \langle t_{\phi z}\rangle^+ \,  \right] \> {\rm d}r
\label{e:Fd}
\end{eqnarray}
The first two terms on the right hand side of this equation describe the
rate at which gravitational binding energy is extracted from matter as it accretes
(i.e. loses angular momentum); the second term in particular describes the rate at
which nonzero MHD stresses at the innermost stable orbit locally dissipate
turbulent energy (in practice, a convenient inner boundary condition is to set this
term equal to zero, although in principle, energy can still be
extracted beyond this boundary). The last term on the right hand side of
(\ref{e:Fd}) describes the rate at which energy is removed through
the disk surface by outflowing  mass and Poynting fluxes.

The result (\ref{e:Fd}) for the radiative flux of an accretion disk
corrected for the effects of outflows can be expressed as
\begin{equation}
F_{\rm d}^+ (r) \approx \frac{3GM\Mdota (r)}{8\pi r^3} \, \left[ \, f_{\rm
    a} (r) - \fw (r) \, \right] \qquad ,
\label{e:Fd1}
\end{equation}
where
\begin{equation}
\fa (r) = \left[ 1 - \frac{\Mdota
    (\ri)}{\Mdota (r)} \left( \frac{\ri}{r} \right)^{1/2} \right] -
\frac{2\pi \ri^2 T_{r\phi} (\ri)}{\Mdota (r) r^2 \Omega}
\label{e:fa}
\end{equation}
is the accretion factor and
\begin{equation}
\fw (r) = \frac{1}{\Mdota (r) r^2 \Omega} \int_{\ri}^{r} \left[ \,  r^2 \Omega \frac{{\rm d}\Mdota}{{\rm d}r}
- 4 \pi r^2  \langle t_{\phi z}\rangle^+ \,  \right] \> {\rm d}r
\label{e:fo}
\end{equation}
is the outflow correction factor (the `$\rm w$' subscript denotes
wind). Note that $\fw$ is equivalent to the fractional  rate of vertical angular
momentum transport in the disk. From
eqns.~(\ref{e:angmom_int}) and (\ref{e:Jdot}), we have
\begin{equation}
\fw(r) = \frac{\dot J_z (r)}{\dot J_{\rm a} (r)} \> .
\end{equation}
Similarly,
\begin{equation}
\fa(r)  = 1 - \frac{\dot J_{\rm a} (\ri) + \dot J_r (\ri)}{\dot  J_{\rm a}(r)} \> ,
\end{equation}
and hence, 
\begin{equation}
\fa(r) - \fw(r) = \frac{\dot J_r (r)}{\dot J_{\rm a}(r)} 
\end{equation}
Therefore, the efficacy of disk radiant emission is entirely determined by the
relative importance of radial transport of angular momentum in the
disk. If vertical transport of angular momentum dominates, then energy
is efficiently removed from the disk before being locally dissipated therein.

It is noteworthy to compare our outflow-modified disk flux with the
disk flux predicted by standard accretion disk theory.
In the absence of outflows
(that is, when ${\rm d}\Mdota /{\rm d} r =0$ and $\langle t_{\phi z}\rangle^+ =0$),
then $\fw = 0$, $\Mdota (\ri) = \Mdota (r) = \Mdot$ and hence, $f_{\rm
  a} = 1 - (\ri /r)^{1/2}$. In the limit of
vanishing stresses at $\ri$, this is equivalent to the small$-r$
correction factor from standard accretion disk theory.\cite{SS73} The radiative flux for a standard disk is then $\Fd^+ (r) \approx 3GM\Mdota/(8\pi r^3)[1 - (\ri/r)^{1/2}]$. That is, standard accretion disk theory predicts that \textit{all} the extracted gravitational binding energy is locally dissipated and then radiated away. In contrast, our model predicts that the local internal dissipation of turbulent stresses is determined by the relative local rates of extraction and removal of gravitational binding energy.
In regions where vertical transport of angular momentum dominates radial
transport, the extracted energy flux is removed from the disk more
efficiently than it is locally dissipated. Thus, a fraction of the
nett accretion power is chanelled into MHD outflows.

\section{An Outflow-Modified Multi-Colour Disk Model}
\label{s:ommcd}

An expression for the {\em total} disk radiative luminosity, $\Ld$, is
obtained by integrating each term in eqn.~(\ref{e:Fd}) over all disk
radii, from $r = \ri$ to $r = \infty$. Similarly, the spectrum of disk
emission can be calculated assuming local blackbody emission and
summing the spectrum from each annulus. This is the Multi-Colour Disk
(MCD) model used for standard disks. \cite{mitsuda84} We will henceforth
refer to our generalized disk prescription
as the Outflow-Modified Multi-Colour Disk (OMMCD) model.
 
Before we can calculate the total luminosity and emission spectrum for
the OMMCD model, it is necessary to define a specific functional form
for the outflows from the disk surface. A simple phenomenological model for
the mass accretion rate is  (see e.g. Refs.~\refcite{wardkon93,li96,cassferr00}
for other self-similar models)
\begin{equation}
\Mdota (r) = \left\{ \begin{array}{ll}
\Mdot \left( \frac{r}{\rw} \right)^p & \> , \> r \leq \rw \\
\Mdot & \> , \> r \geq \rw
\end{array}  \right.
\label{e:Mdota_r}
\end{equation}
where $\rw$ represents a critical wind radius beyond which vertical mass
loss from the disk surface is negligible. The wind mass loss rate then satisfies $\Mdotw
(r) = \Mdot - \Mdota (r)$, from the continuity equation
(\ref{e:Mdot}). We can now use this model to explicitly calculate the
factor $\fa$, which appears as a source term in the disk flux solution
(\ref{e:Fd1}) and which is defined by (\ref{e:fa}). For the
simplest model, we assume that the stresses $T_{r \phi}(\ri)$ at the inner
boundary radius vanish, so that 
\begin{equation}
\fa (r) = \left\{ \begin{array}{ll}
1 - \left( \frac{r}{\ri} \right)^{-(1/2+p)} & \> , \> r \leq \rw \\
1 - \left( \frac{r}{\ri}\right)^{-1/2}
\left( \frac{\rw}{\ri}\right)^{-p}
& \> , \> r \geq \rw
\end{array}  \right.
\label{e:fa_soln}
\end{equation}

We can define a similar model for $\fw (r)$, given by
(\ref{e:fo}). From the
vertically-integrated, differential form of the angular momentum
equation, (\ref{e:angmom}), we have
\begin{equation}
 r^2 \Omega \frac{{\rm d}\Mdota}{{\rm d}r}
- 4\pi r^2 \langle t_{\phi z} \rangle^+ = 
\ddr \left( \Mdota r^2 \Omega \right)
  + \ddr (2\pi r^2 T_{r\phi})   \qquad ,
\label{e:out}
\end{equation}
We wish to define a model in which the relative importance of vertical 
angular momentum transport decreases with increasing radius. We choose a
simple power-law decline, such that
\begin{equation}
 \frac{r^2 \Omega \frac{{\rm d}\Mdota}{{\rm d}r}
- 4\pi r^2 \langle t_{\phi z} \rangle^+} 
{\ddr \left( \Mdota r^2 \Omega \right) } =
\left[ 1 + \frac{\ddr (2\pi r^2 T_{r\phi})}{\ddr (\Mdota r^2 \Omega)}
  \right] = \left( \frac{r}{\ri} \right)^{-w} \qquad ,
\label{e:w}
\end{equation}
with $w > 0$. Thus, in the limit $w \rightarrow 0$, radial
transport of angular momentum by the internal MHD stresses $T_{r\phi}$ is
negligible and the dominant mode of transport is via outflows of
matter and Poynting flux.
Inserting the relation (\ref{e:w}) into (\ref{e:out}) and using
(\ref{e:Mdota_r}) to simplify, we have for $r \leq \rw$
\begin{equation}
r^2 \Omega \frac{{\rm d}\Mdota}{{\rm d}r}
- 4\pi r^2 \langle t_{\phi z} \rangle^+ = 
\left( p+\frac{1}{2} \right) \Mdot c \left( \frac{\rg}{\ri}
\right)^{1/2} \left( \frac{\ri}{\rw} \right)^p \left( \frac{r}{\ri} \right)^{-(1/2-p+w)}
\end{equation}
and for $r \geq \rw$, the corresponding solution is obtained by
setting $p=0$ and $\rm d \Mdota /\rm d r =0$.
This implies the following relation for the disk wind factor 
$\fw$ defined by (\ref{e:fo}):
\begin{equation}
\fw (r) = \left\{ \begin{array}{ll}
\left( \frac{1/2+p}{1/2+p-w} \right) \left( \frac{r}{\ri} \right)^{-w}
\left[ 1 - \left( \frac{r}{\ri} \right)^{-(1/2+p-w)}  \right]
 & \> , \> r \leq \rw \\
\left( \frac{1/2+p}{1/2+p-w} \right)
\left( \frac{r}{\rw} \right)^{-p}
\left( \frac{\rw}{\ri} \right)^{-w}
\left[ 1 - \left( \frac{\rw}{\ri} \right)^{-(1/2+p-w)} \right]  &  \\
\> + \> \left( \frac{1/2}{1/2-w}\right) \left( \frac{r}{\ri} \right)^{-w}
\left[ 1 - \left( \frac{r}{\rw} \right)^{-(1/2-w)} 
\right]
 & , \> r \geq \rw
\end{array}  \right.
\label{e:fw_soln}
\end{equation}
Eqns.~(\ref{e:fa_soln}) and (\ref{e:fw_soln}) for $\fa$ and $\fw$,
respectively, now specify the solution for the OMMCD radiative flux as a function
of radius, given by eqn.~(\ref{e:Fd}).

The total disk luminosity can now be
calculated by integrating the flux over all disk annuli, as follows:
\begin{equation}
\Ld = 2 \int_{\ri}^\infty \Fd (r) \, 2\pi r \, \dr
 = \frac{3}{2} \frac{GM \Mdot}{\ri} \int_{\ri}^\infty \frac{\Mdota
 (r)}{\Mdot} \left( \frac{r}{\ri} \right)^{-2} \left[ \,
 \fa(r) - \fw (r) \, \right]  \, \frac{\dr}{\ri}
\label{e:Ld}
\end{equation}
We can write the solution as
\begin{equation}
\Ld \, = \, \Pa \, - \, \Pw \qquad ,
\end{equation}
where
\begin{equation}
\Pa = \frac{1}{2} \frac{GM \Mdot}{\ri}
\left( \frac{1+2p}{1-p} \right) \left( \frac{\rw}{\ri} \right)^{-p}
\left[ 1 - \frac{3p}{(1+2p)} 
\left( \frac{\rw}{\ri}  \right)^{-(1-p)}
\right]
\label{e:Pa}
\end{equation}
is the total accretion power and
\begin{equation}
\Pw = \frac{1}{2} \frac{GM \Mdot}{\ri}
\left( \frac{1+2p}{1-p+w} \right) \left( \frac{\rw}{\ri} \right)^{-p}
\left[ 1 - \frac{p(3+2w)}{(1+2p)(1+w)}
\left( \frac{\rw}{\ri} \right)^{-(1-p+w)}
\right]
\label{e:Pw}
\end{equation}
is the total power removed by am MHD disk wind. Note that in the limit $p
\rightarrow 0$, corresponding to negligible mass loss (i.e. 
$\Mdota \rightarrow constant$), the accretion power approaches the
solution from standard accretion disk theory, $\Pa = GM\Mdota/2\ri$,
and the wind power is $\Pw \rightarrow \Pa/(1+w)$. Thus, a substantial fraction
of the total accretion power can be removed from the disk by a magnetic torque
alone. This solution represents a Poynting flux dominated outflow and
can be identified with the formation of
relativistic jets that carry away very little matter, but transport a large
amount of mechanical energy very efficiently across vast distances.

The emission spectrum predicted by the OMMCD model can be calculated in
the same way as that for an MCD model (i.e. a standard accretion
disk). Assuming each annulus in the disk radiates locally like a
blackbody, $B_\nu$, the total disk spectrum is calculated by summing
the individual spectra from each annulus:
\begin{equation}
L_{\rm d , \nu} = 2 \int_{\ri}^\infty \pi B_\nu [ T(r) ] \, 2\pi r \,
\dr
\qquad ,
\end{equation}
where $T(r) = [ \Fd (r) /\sigma ]^{1/4}$ is the effective disk
temperature of each annulus and $\sigma$ is the Stefan-Boltzmann constant.

\begin{figure}[th]
\centerline{\includegraphics[width=3.5in]{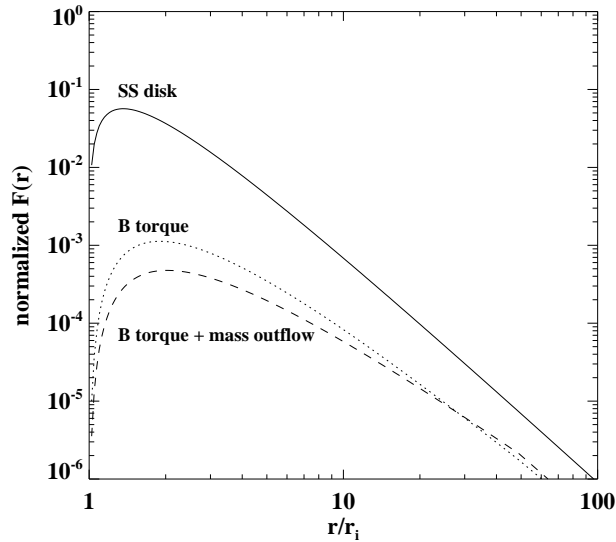}}
\caption{Predicted radiative energy flux emitted by an accretion disk
  as a function of radius $r$, in units of $3GM\Mdota (r)/(8\pi r^3)$,
  where $\Mdota (r)$ is the local mass accretion rate
  (c.f. eqn.~\ref{e:Fd1}). The solid line show the total radiative
  flux profile for a standard Shakura-Sunyaev (SS) disk. The dotted
  line indicates the radial flux profile for a disk modified by a
  magnetized outflow (with $w=0.1$) with negligible mass loss
  ($p=0.1$). The dashed line shows the radial flux profile for a disk
  modified by both a magnetic torque (with $w=0.1$) and a mass-loaded
  outflow (with $p=0.5$). Mass loss is confined to radii within
  $50\ri$ (see Sec.~\ref{s:ommcd} for definitions of the model parameters). 
\protect\label{fr}}
\end{figure}


\begin{figure}[ht]
\centerline{\includegraphics[width=3.5in]{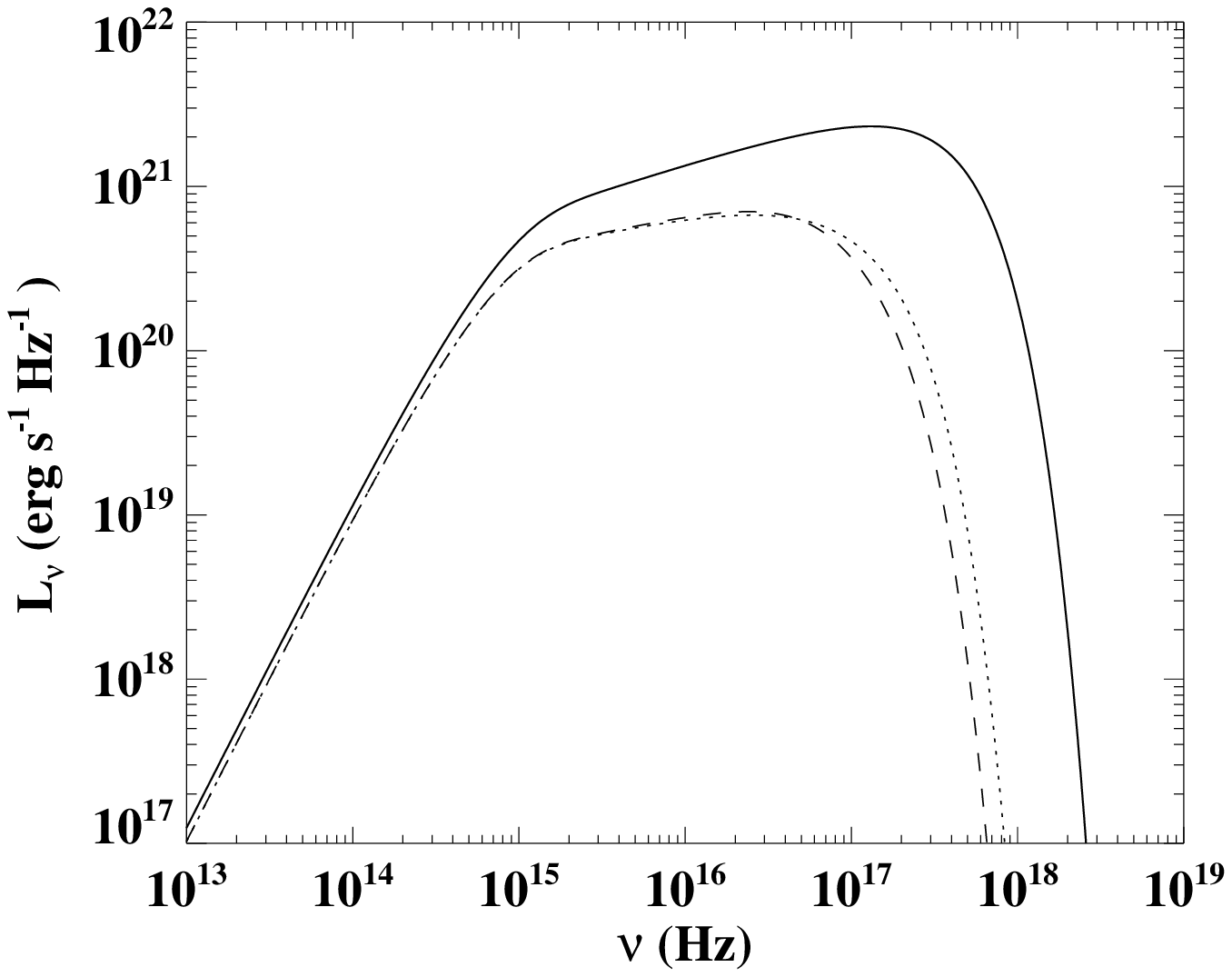}}
\centerline{\includegraphics[width=3.5in]{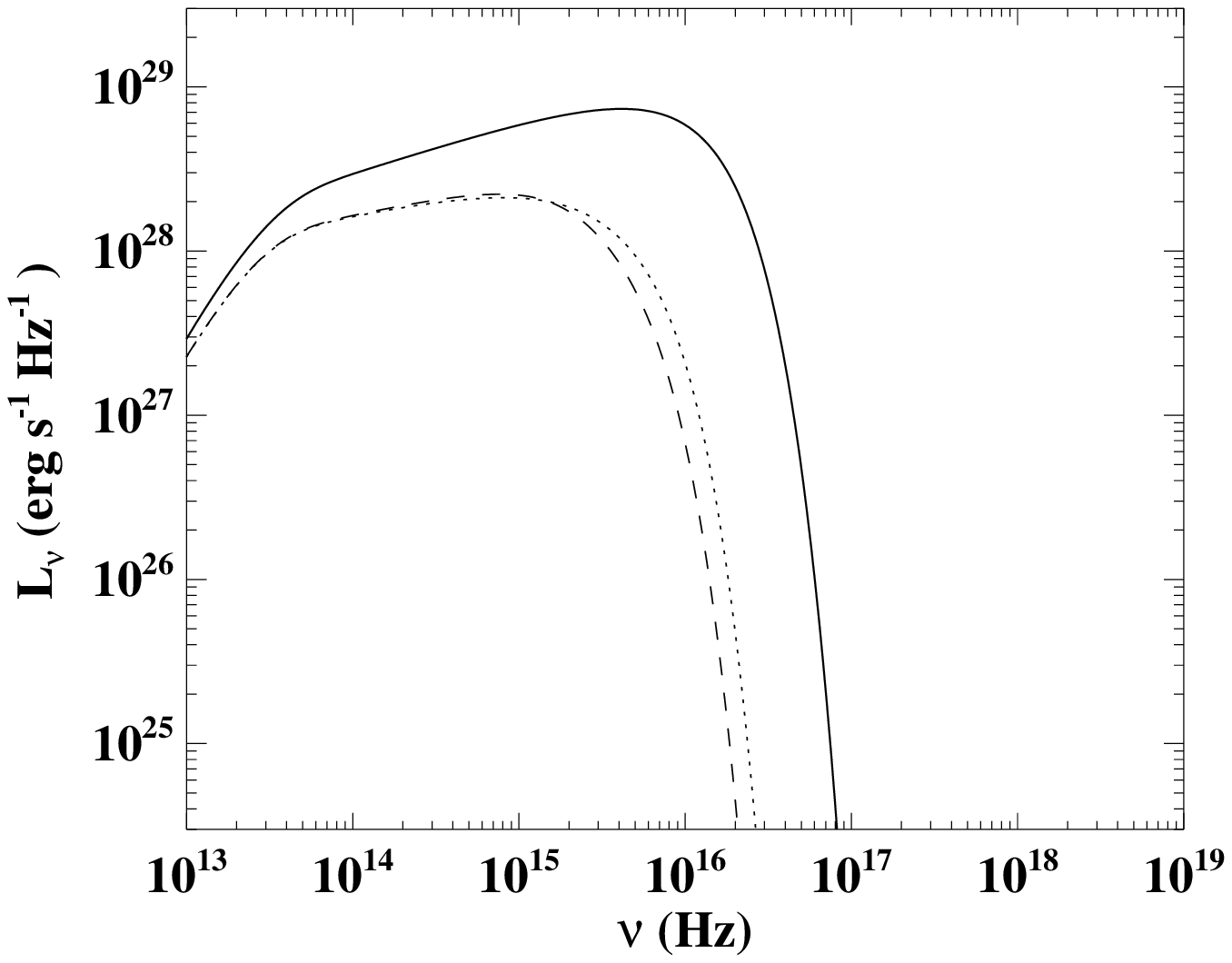}}
\caption{Top: Predicted emission spectra for an accretion disk around a
  $M = 10\,M_\odot$ black hole accreting at a rate of $\Mdot = 2.7
  \times 10^{-7} \, M_\odot \, {\rm yr}^{-1}$ at $r = \infty$. Bottom:
  Predicted spectra for $M = 10^7 M_\odot$ and $\Mdot = 0.27 \,
  M_\odot \, {\rm yr}^{-1}$ at $r = \infty$. In both cases, solid
  lines show the
  spectra for a standard Shakura-Sunyaev (SS) disk. Dotted
  lines indicate the spectra for a disk modified by a
  magnetic torque, with $w=0.1$ and with negligible mass loss
  ($p=0.1$). Dashed lines indicate the
  spectra for a disk modified by both a magnetic torque (with $w=0.1$)
  and a mass-loaded outflow, with $p=0.5$. Mass loss is restricted to
  radii within $50\ri$ (see Sec.~\ref{s:ommcd} for
  definitions of these parameters and Fig.~\ref{fr} for the
  corresponding local radiative fluxes).
\protect\label{Lnu}}
\end{figure}

Figure~\ref{fr} shows plots of the disk flux $\Fd (r)$  for three
different models of disk accretion around a
black hole:  1. the standard ``Shakura-Sunyaev''\cite{SS73} (MCD)
model; 2. the OMMCD model with negligible
mass outflow ($p=0.1$) and a magnetic torque carrying away $\approx
90\,$\% of the accretion power ($w=0.1$); and 3. an OMMCD model with a
stronger mass outflow (with $p=0.5$) as well as a magnetic torque
(with $w=0.1$). The mechanical energy carried away by this outflow is
also $\approx 90\,$\% of the total accretion power. In all cases, mass
loss is restricted to radii $r \leq 50\ri$. Model 2 corresponds to a
disk modified by a Poynting flux dominated outflow such as a
magnetized corona and/or jet. In model 3, $\Pa$ is lower (by about
a factor of 2) than that in the first two models because there is less
matter reaching $\ri$.

The corresponding disk emission spectra $L_{\rm d , \nu}$ are shown in
Figure~\ref{Lnu} for a stellar-mass black hole with $M = 10 M_\odot$
and accretion rate $\Mdot = 2.7 \times 10^{-7} \, M_\odot \, {\rm
  yr}^{-1}$ at $r = \infty$ and for a supermassive black hole with $M
= 10^7 M_\odot$ and $\Mdot = 0.27 \, M_\odot \, {\rm yr}^{-1}$. These
accretion rates represent the nett mass influxes at $r=\infty$ and
correspond to the Eddington rates when the radiative efficiency is
$\eta =0.1$. This is strictly only the case for the SS disk model. When outflows are important, either due to a magnetic torque alone or in addition to a mass-loaded disk wind, the radiative efficiency can fall below the nominal $10$\% predicted by standard theory because the outflows remove energy from the disk (as well as angular momentum), so there is relatively less energy to dissipate and radiate away.

As is clearly evident in Figs.~\ref{fr} and \ref{Lnu}, outflows can substantially modify the local disk emission profile and hence, the overall spectrum of emission. The generalized OMMCD model
predicts a high-energy cut-off in the emission spectrum at lower energies than that predicted by the standard SS model. This results from the importance of outflows at small radii, where the highest-energy
emission originates. In the case of stellar-mass black holes, the spectrum is most affected at X-ray energies (between $10^{17-18}\,{\rm Hz}$), while for supermassive black holes, it is the extreme ultraviolet region ($\sim 10^{16}\,{\rm Hz}$) that is most affected by outflows. The OMMCD model also predicts a broadband region of the disk spectrum (around visual frequencies) that is much flatter than the characteristic $\nu^{1/3}$ law predicted by standard theory.

Our results demonstrate that the modifications to a standard disk spectrum as a result of energetic outflow phenomena are clearly not negligible when the outflows are primarily responsible for the removal of angular momentum from accreting matter. Direct observational verification of the predicted OMMCD spectrum may be possible for only some types of accreting sources, however, as the outflows will also produce their own characteristic radiative signatures that may overlap in the spectral energy band where we predict the disk to be most strongly modified. This can confuse our interpretation of the emergent observed spectrum.

In the case of galactic X-ray binaries (XRBs), for instance, the disk spectrum can be substantially modified at X-ray energies (typically above $1$\,keV, corresponding to frequencies above $10^{17}$\,Hz -- see Fig.~\ref{Lnu}). These sources often exhibit a power-law X-ray spectrum above $1$\,keV that is thought to arise in a magnetized, tenuous atmosphere (a "corona") and/or relativistic jet as a result of inverse Compton scattering off disk photons (see Ref.~\refcite{mcclinrem06} for a review). Existing models for this emission process simply use a standard disk spectrum for the seed photon distribution. However, this is clearly not a self-consistent calculation since the formation of a corona and/or jet from disk magnetic fields will inevitably modify the disk radial structure and hence,  photon spectrum, as we have shown here. Thus, although we would not expect to directly see the modified disk spectrum in XRBs, our model can be indirectly tested by using the underlying OMMCD spectrum as the source of seed photons that are upscattered in a disk corona and/or jet. By comparing the predicted and observed X-ray spectra, we can improve our interpretation of the source characteristics and obtain tighter constraints on key physical parameters such as mass accretion rate.

The above type of calculation has been performed by us for ultra-luminous X-ray sources (ULXs). \cite{freeland06} These are exceptionally luminous XRBs found in external galaxies. They probably involve accretion onto a black hole more massive than those found in galactic XRBs. They provide an effective test for our modified disk model because some ULXs exhibit a low energy ($\ltapprox 1\,{\rm keV}$) spectral component that can be interpreted as disk emission. However, because these sources emit virtually all of their radiative power as hard, power-law X-rays, which must necessarily be produced outside the disk, the accretion disk must be substantially modified and a standard disk model should not be used to fit the soft spectral component. Indeed, we show \cite{kuncic06} that the OMMCD spectrum can adequately fit the observed soft spectral component, implying a black hole mass $M \sim 100 M_\odot$.

A similar test can also be performed for quasars and other AGN, which are powered by accretion onto a supermassive ($10^{6-9}\,M_\odot$) black hole. These sources exhibit a prominent optical-ultraviolet (UV) continuum feature known as the "big blue bump". \cite{sanders89} This energetically significant  feature is generally interpreted as accretion disk emission. \cite{shields78,malksarg82} However, attempts at fitting accretion disk spectral models to the observed spectra have had mixed success
(see Ref.~\refcite{koratblaes99} for a review). For AGN, the disk
spectrum will be most strongly modified by outflows at ultraviolet
(UV) wavelengths (see Fig.~\ref{Lnu}). Although there is less overlap
between the modified disk and corona/jet X-ray emission than in the
XRB case, the poor transmission of UV radiation through Earth's
atmosphere means that direct observations of the predicted OMMCD
spectrum in AGN are not straightforward. Nevertheless, spectral energy
distributions have been compiled from optical observations of
high-redshift quasars and from satellite UV observations with the
\textit{Hubble Space Telescope (HST)} and the \textit{Far Ultraviolet
  Spectroscopic Explorer (FUSE)}. Interestingly, these observations
have revealed a far-UV break in the spectral energy distributions of
quasars (see Ref.~\refcite{trammell07} and references cited
therein). Specifically, the observed spectra decline dramatically at
wavelengths shorter than $1100$\,\AA\, (i.e. frequencies above $3
\times 10^{15}\,{\rm Hz}$). This is difficult to explain with standard
accretion disk models, which predict that the spectrum should continue
to rise into the extreme UV (see Fig.~\ref{Lnu}), but is consistent
with the predictions our modified disk model (see Ref.~\refcite{kunbick07}).

\section{Concluding Remarks}

The standard theory for astrophysical disk accretion has enjoyed
remarkable success since its inception over 30 years ago. However,
ongoing rapid advances in instrumental technology are placing
increasingly tighter observational constraints on theoretical models
that use a standard accretion disk. The ubiquity of energetic outflow
phenomena in accreting astrophysical sources gives us two important
clues to the underlying processes responsible for accretion:
1. vertical transport of angular momentum cannot be ignored and may
indeed be the dominant mode of transport at small radii; and 2. the
removal of accretion energy from the disk  must result in a modified
radial disk structure, particularly at small radii, and consequently, a
disk spectrum that is dimmer and redder than a standard disk accreting at the same rate. We have presented a simple model that can be used
to quantitatively calculate model spectra for disks modified by a
magnetized jet and/or corona and by mass-loaded winds. These model
spectra can be applied to a variety of different sources to improve
our interpetation of the observations and more accurately determine
key physical parameters such as mass accretion rate and black hole
mass, as well as the partitioning of accretion power into radiative
and non-radiative forms.

\section*{Acknowledgments}

ZK acknowledges support for this work from a University of Sydney R\&D grant.




\end{document}